\def\Granadadep{Departamento de F\'\i sica Te\'orica y del Cosmos, Facultad
de Ciencias, Universidad de Granada, Campus de Fuentenueva, Granada 18002,
Spain}

\def\Granadainst{Instituto de F\'\i sica Te\'orica y Computacional Carlos I, 
Facultad
de Ciencias, Universidad de Granada, Campus de Fuentenueva, Granada 18002, 
Spain}

\def\Valencia{IFIC, Centro Mixto Universidad de Valencia-CSIC, Burjasot
              46100-Valencia, Spain.}
\def\Comision{Work partially supported by the DGICYT.}

\def\um{\f{1}{2}}
\def\p{\partial}
\def\TG{\tilde{G}}
\def\tg{\tilde{g}}

\def\hs{\bar{s}}
\def\s{s}

\def\hz{\bar{z}}

\def\hm{{\bar{m}}}
\def\htheta{{\bar{\theta}}}
\def\f{\frac}
\def\l[{\left[}
\def\r]{\right]}

\def\TXL{{\tilde{X}}^L}
\def\TXR{{\tilde{X}}^R}
\def\L{L}

\def\heta{\bar{\eta}}
\def\hc{\bar{\alpha}}
\def\hcc{{\bar{\alpha}^*}}
\def\b{b}
\def\bb{{\bf b}}
\def\U{U}
\def\hU{\bar{U}}
\def\a{{\hat{a}}}
\def\ac{{\hat{a}}^{\dag}}
\def\os{a}
\def\osc{a^*}
\def\bos{{\bf a}}
\def\bosc{{\bf a}^*}
\def\bx{{\bf x}}
\def\bp{{\bf p}}
\def\bn{{\bf n}}
\def\bk{{\bf k}}
\def\cc{\alpha^*}
\def\c{\alpha}

\def\hchi{\bar{\chi}}

\def\nn{\nonumber}
\def\ni{\noindent}

\def\be{\begin{equation}}
\def\ee{\end{equation}}
\def\bea{\begin{eqnarray}}
\def\eea{\end{eqnarray}}
\def\ba{\begin{array}}
\def\ea{\end{array}}
\def\hej{\hat{{\bf e}}_j}
\def\heu{\hat{{\bf e}}_1}
\def\hed{\hat{{\bf e}}_2}
\def\het{\hat{{\bf e}}_3}
\def\A{L}

\newcommand{\deriv}[1]{ \frac{\partial}{\partial #1} }

\newcommand{\xl}[1]{ {\tilde{X}}^{L}_{#1} }

\newcommand{\xr}[1]{ {\tilde{X}}^{R}_{#1} }

\documentstyle[11pt]{article}

\begin{document}
 


\hbox{ }

\vskip 3 cm

\begin{center} 
{\LARGE {\bf Generalized  Conformal Symmetry and Extended Objects 
 from the Free Particle$^1$ }}
\end{center}

\bigskip
\bigskip

\centerline{ M. Calixto$^{2,4}$, V. Aldaya$^{2,3}$  
       and J. Guerrero$^{2,4}$ }  

\bigskip

\footnotetext[1]{\Comision}
\footnotetext[2]{\Granadainst} \footnotetext[3]{\Valencia}
\footnotetext[4]{\Granadadep}  

\bigskip

\begin{center}
{\bf Abstract}
\end{center}

\small

\begin{list}{}{\setlength{\leftmargin}{3pc}\setlength{\rightmargin}{3pc}}
\item The algebra of linear and quadratic function of basic observables on 
the phase space of either the free particle or the harmonic oscillator 
possesses a finite-dimensional anomaly. The quantization of these systems
outside the critical values of the anomaly leads to a new degree of 
freedom  which shares its internal character with spin, but nevertheless 
features an infinite number of different states. Both 
are associated with the transformation properties of wave 
functions under the Weyl-symplectic group $WSp(6,\Re)$. The physical 
meaning of this new degree of freedom can be established, with a major 
scope, only by analysing the quantization of 
an infinite-dimensional algebra of diffeomorphisms generalizing string 
symmetry and leading to more general extended objects.
 
\end{list}

\normalsize

\vskip 1cm

\section{Introduction}

The irreducible representations of the Schr\"odinger group were first 
studied by Niederer \cite{Niederer} and Perroud \cite{Perroud}, who found 
for some of these an unclear association with the concept of elementary
systems, as a consequence of the appearance of an infinite number of 
internal states. These internal states correspond to the infinite-dimensional 
carrier space that supports the irreducible representations with Bargmann 
index $k$ \cite{Bargmann} of the (non-compact) $SL(2,\Re)$ 
subgroup  replacing the time translation, in either the Galilei or the 
harmonic oscillator (Newton) group, to give the Schr\"odinger symmetry. 

Representations of the Schr\"odinger group, or more generally $WSp(6,\Re)$, 
with non-trivial 
$k$ should correspond to linear systems (free particle, harmonic oscillator, 
etc.) with a new internal degree of 
freedom, much in the same way representations of the (compact) $SO(3)$ subgroup
with non-trivial index $s$ are associated with elementary particles with
spin degree of freedom, although in this case supporting a finite number of 
states 
($2s+1$ indeed). However, unlike spin, which occurs in nature for 
any half-integer value of $s$, only the value $k=\frac{d}{4}$ ($d$ stands for 
the spatial dimension)  is currently found in physical systems such as 
those described in Quantum Optics \cite{QO}. 

The differences and analogies between $k$ and $s$ are clarified if we realize
that the Schr\"odinger group is an anomalous symmetry \cite{Anomalias} and that
$k=\frac{d}{4}$ is the quantum value of the anomaly. This means that 
$k=\frac{d}{4}$ is the value that quantization associates with the na\"\i vely
expected, classical value $k=0$ corresponding to a classical system without
any extra internal degree of freedom (hereafter we shall call $\hs=-k$ the 
{\it symplectic spin} of just the {\it symplin} for the sake of brevity). 
Thus, roughly speaking we can say 
that this value of $k$ corresponds to a quantum system without symplin.

However, the possibility of a non-trivial value of $k$, even 
though anomalous, prompts us to wonder whether or not quantizing a free 
system for a non-critical value of $k$ makes any sense. In this paper,
we provide a detailed quantization of a linear system (three-dimensional 
isotropic harmonic oscillator) bearing both spin and 
symplin, seeking tp clarify the behaviour of the Hilbert space of the 
theory according to the values of $k$, in full agreement with the behaviour
of more standard, infinite-dimensional, anomalous theories.

Unlike the spin, the physical meaning of which seems to be well 
described in terms of fermionic and bosonic objects, the symplin does 
not appear to fit any known characteristic of the 
elementary particle. Rather, it seems to be understood  as forming 
part of a larger set of degrees of freedom originating in the free 
particle and conforming an extended object which proves to 
generalize other physical systems bearing conformal symmetry. 
More precisely, this extended object arises when  trying to quantize
more classical observables than those allowed by the well-known 
{\it no-go} theorems \cite{no-go}, the Schr\"odinger anomaly being the 
first obstruction to standard quantization. In fact, a proper choice of 
an infinite-dimensional basis of the classical Poisson algebra on the 
phase space of a linear system (the free particle or the harmonic 
oscillator) and an adequate treatment of Lie-algebra 
cohomology and central extensions, as well as 
anomalies, will lead to a dynamical system which should be understood as an 
extended object with an infinite number of degrees of freedom. This symmetry, 
which in particular contains string symmetry, has the Schr\"odinger 
algebra as the maximal finite-dimensional subalgebra. In this way, 
the particle with  symplin mentioned above would appear as 
the simplest (and the only finite-dimensional) part of this generalized 
object, yet possessing one of these extended degrees of freedom.

This paper is organized as follows. Sec. II is devoted to presenting the 
Group Approach to Quantization (GAQ) technique, which we shall use 
extensively through the paper. This formalism proves 
especially suitable in discussing the connection between dynamical degrees 
of freedom and group cohomology and, therefore, the role played by 
anomalies. In Sec. III, we compute and fully describe the
irreducible representations of the Schr\"odinger group with non-trivial 
indices $s$ and $k$, as corresponding to a particle with spin and symplin.
In particular, we dwell on the specific situation $k=\frac{d}{4}$, 
for which the Hilbert space acquires an exceptional reduction, reflecting 
the presence of the anomaly; we compare this situation with that of ordinary 
bosonic 
string, which is better known. Finally, in Sec. IV, and to best understand the 
meaning of the extra degree of freedom introduced here, we pursue in depth 
the ideas which have lead us to the $k$ degree of freedom. In so doing, we 
seek to quantize the Poisson algebra associated with a free particle far 
beyond no-go theorems and na\"\i ve analyticity obstructions. We end with a 
generalization of the DeWitt $w_\infty$ algebra, which contains, in 
particular, the full Virasoro algebra $\{L_n, n\in Z\}$ and the algebra of 
string modes $\{\alpha_m\}$, closing a semi-direct product. The actual 
quantization of the entire algebra and the analysis of the possible 
anomalies is only sketched. 
 
\section{Quantization on a group, (pseudo-)co-homology and anomalies}

The starting point of GAQ \cite{GAQ} is a group $\TG$ (the quantization group) 
with a principal fiber bundle 
structure $\TG(M,T)$, having $T$ as the structure group and $M$  
the base. The group $T$ generalizes 
the phase invariance of Quantum Mechanics. Although the situation can be 
even more general \cite{Ramirez}, we shall
start with the rather general case in which $\TG$ is a central extension of
a group $G$ by $T$ [$T=U(1)$ or even $T={\bf C}^*=\Re^+\otimes U(1)$]. 
For the one-parameter group 
$T=U(1)$, the group law for $\TG=\{\tilde{g}=(g,\zeta)/ g\in G, 
\zeta\in U(1)\}$ adopts the following 
form:
\be
\tilde{g}'*\tilde{g}=(g'*g,\zeta'\zeta e^{i\xi(g',g)})
\ee
\ni where $g''=g'*g$ is the group operation in $G$ and $\xi(g',g)$ is a two 
cocycle of $G$ on $\Re$ fulfilling:
\be
\xi(g_2,g_1)+\xi(g_2*g_1,g_3)=\xi(g_2,g_1*g_3)+\xi(g_1,g_3)\;\;, g_i\in G. 
\ee
\ni In the general theory of central extensions \cite{Extensiones}, two 
two-cocycles are said to be equivalent if they differ in a coboundary, i.e. a 
cocycle which can be written in the form  $\xi(g',g)=\delta(g'*g)-
\delta(g')-\delta(g)$, where $\delta(g)$ is 
called the generating function of the coboundary. However, although cocycles 
differing on a coboundary lead to equivalent central extensions as such, 
there are some coboundaries which provide a non-trivial connection on the 
fibre bundle $\TG$ and Lie-algebra structure constants different from that 
of the direct product $G\otimes U(1)$. These are generated by a function 
$\delta$ with a non-trivial gradient at the 
identity, and can be divided into equivalence Pseudo-cohomology subclasses: 
two pseudo-cocycles are equivalent if they differ in a coboundary generated 
by a function with trivial gradient at the identity \cite{Saletan,Pseudoco,
Marmo}. Pseudo-cohomology plays an important role in the theory of 
finite-dimensional semi-simple group, as they have trivial cohomology. For 
them, Pseudo-cohomology classes are associated with coadjoint orbits 
\cite{Marmo}.

The right and left finite actions of the group $\TG$ on itself provide two 
sets 
of mutually commuting (left- and right-, respectively) invariant vector fields:
\be
\TXL_{\tg^i}=\f{\p {\tg''}{}^j}{\p \tg^i}\mid_{\tg=e}\f{\p}{\p \tg^j},\;\;\;
\TXR_{\tg^i}=
\f{\p {\tg''}{}^j}{\p {\tg'}{}^i}\mid_{\tg'=e}\f{\p}{\p \tg^j},\;\;\; 
\l[\TXL_{\tg^i},\TXR_{\tg^j}\r]=0,\label{camposlr}
\ee
\ni where $\{\tg^j\}$ is a parameterization of $\TG$.
The GAQ program continues finding the left-invariant 1-form $\Theta$ (the 
{\it Quantization 1-form})
associated with the central generator $\TXL_\zeta=\TXR_\zeta, \zeta\in T$, 
that is, the $T$-component
$\tilde{\theta}^{L(\zeta)}$ of the canonical left-invariant 1-form
$\tilde{\theta}^L$ on $\tilde{G}$. This constitutes the generalization of the 
Poincar\'e-Cartan form of Classical Mechanics (see \cite{Abraham}).
The differential $d\Theta$ is a
{\it presymplectic} form and its {\it characteristic module}, 
$\hbox{Ker}\Theta\cap\hbox{Ker} {}d\Theta$, is generated by a 
left subalgebra ${\cal G}_\Theta$ called {\it
characteristic subalgebra}. The quotient $(\TG, \Theta)/{\cal G}_\Theta$ is a
{\it quantum manifold} in the sense of Geometric Quantization 
\cite{GQ1,GQ2,GQ3,GQ4}. The trajectories generated by the vector fields  in 
${\cal G}_\Theta$ constitute the generalized equations of motion of 
the theory (temporal evolution, rotations, gauge transformations, etc...), 
and the ``Noether" 
invariants under those equations are $F_{\tg^j}\equiv i_{\TXR_{\tg^j}}\Theta$, 
that is, the contraction of right-invariant vector fields with 
the Quantization 1-form. 
Those vector fields with null Noether charge are called 
{\it gauge} \cite{config2} and the subspace expanded by all 
the gauge vector fields is termed {\it gauge subalgebra}, 
which proves to be an ideal of the whole algebra of $\TG$.  

Let ${\cal B}(\TG)$ be the set of complex-valued
$T$-{\it functions} on $\TG$ in the sense
of principal bundle theory: 
\be
\psi(\zeta*\tg)=D_{T}(\zeta)\psi(\tg), \;\;\zeta\in T\label{tcondition}
\ee
\ni where $D_{T}$ is the natural representation of $T$ on the complex numbers 
${\bf C}$. The representation of $\TG$ on ${\cal B}(\TG)$ generated
by ${\tilde{\cal G}}^R=\{\TXR\}$ is called {\it Bohr Quantization} and is 
{\it reducible}. The true {\it Quantization} is achieved when this 
pre-quantization
is fully reduced, usually by means of the 
restrictions imposed by a {\it full polarization} ${\cal P}$:
\be
\TXL\psi_p=0,\;\;\forall \TXL\in {\cal P}\label{defpol}
\ee
\ni which is a maximal, horizontal (i.e. in $Ker\Theta$) left
subalgebra of ${\tilde{\cal G}}^L$ which contains ${\cal G}_\Theta$.
It should be noted, however, that the existence of a full polarization, 
containing the whole subalgebra ${\cal G}_\Theta$, is not guaranteed. 
In case of such a breakdown, called {\it anomaly}, or simply by the desire of 
choosing a preferred representation space, a {\it higher-order polarization} 
must be imposed \cite{Anomalias,Marmo,Napoles}. A higher-order 
polarization is a maximal subalgebra of the left enveloping 
algebra $U\tilde{{\cal G}}^L$ with no intersection with the abelian 
subalgebra of powers of $\xr{\zeta}$.
   
The group $\TG$ is irreducibly represented on the space 
${\cal H}(\TG)\equiv\left\{|\psi\rangle \right\}$ of (higher-order) 
polarized wave functions. If we denote by 
\be
\psi_p(\tg)\equiv\langle \tg_p|\psi\rangle \,,\,\,
\psi'{}^*_p(\tg)\equiv\langle \psi'|\tg_p\rangle 
\ee
\ni the coordinates of the ``ket" $|\psi\rangle $ and the ``bra" 
$\langle \psi|$ in a representation defined through 
a polarization ${\cal P}$ (first or higher order), then, a scalar product 
on ${\cal H}(\TG)$ can be naturally defined as:

\be
\langle \psi'|\psi\rangle \equiv
\int_{\TG}{ v(\tg)\psi_p'{}^*(\tg)\psi_p(\tg)},\label{scalarp}
\ee
\ni where  
\be
 v(\tg)\equiv\theta^L_{\tg^i}\wedge\stackrel{\hbox{dim}(\TG)}{...}
\wedge\theta^L_{\tg^n}\label{volumen}
\ee
\ni is the left-invariant integration volume in $\TG$ and
\be
1=\int_{\TG}{|\tg_p\rangle  v(\tg)\langle \tg_p|}\label{closure}
\ee
\ni formally represents a {\it closure} relation.
A direct computation proves that, with this scalar product, 
the group $\TG$ is unitarily represented through the 
{\it left} finite action ($\rho$ denotes the representation)
\be
\langle \tg_p|\rho(\tg')|\psi\rangle \equiv\psi_p(\tg'{}^{-1}*\tg) 
\label{leftaction}
\ee

{\it Constraints} are consistently incorporated into the theory by 
enlarging  the structure group $T$ (which always includes $U(1)$), i.e, 
through  $T$-function conditions:
\be
\rho(\tilde{t})|\psi\rangle =D^{(\epsilon)}_T(\tilde{t})|\psi\rangle\,,\;\;
\tilde{t}\in T 
\ee
\ni or, for continuous transformations,
\be
\TXR_{\tilde{t}}|\psi\rangle =dD^{(\epsilon)}_T(\tilde{t})|\psi\rangle \;, 
\label{const}
\ee 
$D^{(\epsilon)}_T$ means a specific representation of $T$  [the index 
$\epsilon$ parametrizes different (inequivalent) quantizations] and 
$dD^{(\epsilon)}_T$ is its differential.

 Ofcourse, for a non-central structure 
group $T$, not all the right operators $\TXR_{\tg}$ will preserve 
these constraints; a sufficient condition for a subgroup  $\TG_T\subset\TG$ 
to preserve the constraints is (see \cite{FracHall,Medular}):
\be
 \l[\TG_T,T\r]\subset \hbox{Ker}D^{(\epsilon)}_T\label{good}
\ee
\ni  [note that, for the trivial representation of $T$, the subgroup $\TG_T$ 
is in fact the {\it normalizer} of $T$]. 
$\TG_T$ takes part of the set of {\it good} operators \cite{Ramirez}, of
the enveloping algebra $U\tilde{{\cal G}}^R$ in general, for which 
the subgroup $T$ behaves as a {\it gauge} group (see \cite{config2} for a 
thorough study of gauge symmetries and constraints from the point of 
view of GAQ). A more general situation can be posed when the 
constraints are lifted to the higher-order level, not necessarily first 
order as in (\ref{const}); that is, they are a subalgebra of the right 
enveloping 
algebra $U\tilde{{\cal G}}^R$. A good example of this last case 
is found when one selects representations labelled by a value $\epsilon$ 
of some Casimir operator $C$ of a subgroup $\TG_C$ of $\TG$ \cite{Conforme}. 

In the more general case in which $T$ is not a trivial central extension,
$T \neq \check{T} \times U(1)$, where $\check{T} \equiv T/U(1)$ -i.e. $T$ 
contains second-class constraints- the conditions (\ref{const}) are not all
compatible and we must select a subgroup $T_B = T_p \times U(1)$, where
$T_p$ is the subgroup associated with a right polarization subalgebra of the
central extension $T$ (see \cite{Ramirez}).

For simplicity, we have sometimes made use of infinitesimal (geometrical) 
concepts, but all this language can be translated to their finite (algebraic) 
counterparts (see \cite{Ramirez}), a desirable way of proceeding when 
discrete  
transformations are incorporated into the theory.

Before ending this section, we eish to insist a bit more on the concept of 
(algebraic) anomaly, which will be quite relevant to what follows. We have 
introduced  the concept of full polarization 
subalgebra intended to reduce the representation obtained through the
right-invariant vector fields acting on $T$-equivariant functions on the 
group. It contains ``half'' of the symplectic vector fields as well as the 
entire characteristic subalgebra. If the full reduction is achieved, 
the whole set of physical operators can be rewritten in terms of the basic 
ones, i.e. those for which the left counterpart is not in the characteristic 
subalgebra ${\cal G}_\Theta$. For instance, the 
energy operator for the free particle can be written as 
$\frac{\hat{p}^2}{2m}$, the angular momentum in 3+1 dimensions is the 
vector product $ \hat{\bx}\times  \hat{\bp}$, or the energy for 
the harmonic oscillator is $\hat{a}^{\dag}\hat{a}$ (note that, since we 
are using first-order polarizations, all these operators are really written
as first-order differential operators).

However, the existence of a full polarization is guaranteed
only for semisimple and solvable groups \cite{Kirillov}.
We define  an {\it anomalous} group \cite{Anomalias,Marmo}  as  
a central extension 
$\TG$ which does not admit any full polarization 
for certain values of the (pseudo-)cohomology parameters, called the 
{\it classical} values of the anomaly (they are called classical because 
they are associated with the coadjoints orbits of the group $\TG$, that is, 
with the classical phase space of the physical system). 
Anomalous groups feature another set of values of the (pseudo-)cohomology 
parameters, called the {\it quantum} values of the anomaly, for which the 
carrier space associated 
with a full polarization contains an invariant subspace.
For the classical values of the anomaly, the classical solution manifold 
undergoes a reduction in dimension, thus increasing the number of 
(non-linear) relationships among Noether invariants,  whereas for the quantum 
values the number of 
basic operators decreases on the invariant (reduced) subspace due to
the appearance of (higher-order) relationships among the quantum operators.

We should comment that the anomalies we are dealing with in this paper are
of {\it algebraic} character in the sense that they appear at the Lie algebra 
level,
and must be distinguished from the {\it topologic anomalies} which are
associated with the non-trivial homotopy of the (reduced) phase space 
\cite{FracHall}.

The non-existence of a full  polarization is traced back to 
the presence in the characteristic subalgebra, for certain values 
of the (pseudo-)cohomology parameters (the classical values of the anomaly), 
of some elements the adjoint action of which is not diagonalizable in the 
``$x-p$-like" algebra subspace. The anomaly problem presented here parallels 
that of the non-existence of invariant polarizations in the Kirillov-Kostant 
co-adjoint
orbits method \cite{Gotay}, and the conventional anomaly problem in
Quantum Field Theory which manifests itself through the appearance of
central charges in the quantum current algebra, absent from the classical
(Poisson bracket) algebra \cite{Jackiw}.
 
The practical way in which an anomaly appears and how a higher-order 
polarization
fully reduces the Hilbert space of the quantum theory for the particular
quantum value of the anomaly will be apparent with the finite-dimensional 
example
discussed in the next section and the comparison with other much better 
known infinite-dimensional cases.

\section{Internal degrees of freedom associated with the elementary particle}

Internal degrees of freedom of a linear, quantum system 
with $\Re^{2d}$ as phase space are generally associated with non-trivial 
transformation properties  of the phase $\zeta$ of the wave function  under 
the symplectic group $Sp(2d,\Re)\subset WSp(2d,\Re)$. Their presence
is evident in the 
emergence of central charges in the Lie algebra $sp(2d,\Re)$ of symplectic 
transformations --which is isomorphic to the classical Poisson algebra of 
all quadratic functions of the position $x_j$ and the conjugate 
momentum $p_j$-- and their origin is cohomological. The simplest case to be 
considered is $d=1$ (particle on a line) for which the Lie algebra of the 
symplectic group  
$Sp(2,\Re)\simeq SL(2,\Re)\simeq SU(1,1)$, isomorphic to 
the Poisson algebra generated by $\{\f{1}{2}x^2, \f{1}{2}p^2, xp\}$, appears 
to be naturally extended, providing a representation with Bargmann index 
$k=\f{1}{4}$. As already mentioned, there is no {\it a priori} physical 
significance for other representations carrying different values of the 
Bargmann index $k$ (the symplin $\hs$ for us). 
To construct explicitly these representations 
and compare them with the most usual case of the spin, we shall consider the 
$d=3$ case and restrict ourselves to the Schr\"odinger subgroup of the 
Weyl-symplectic $WSp(6,\Re)$ group, where we have replaced the $Sp(6,\Re)$ 
group by its $SL(2,\Re)\otimes SO(3)$ subgroup, for which the Lie algebra is 
isomorphic to the Poisson algebra generated by 
$\{\f{1}{2}{\bx}^2,\, \f{1}{2}{\bp}^2,\,{\bx\cdot \bp};\,{\bx\times \bp}\}$. 
At this juncture, it will be convenient 
to use  a oscillator-like parametrization in terms of the usual 
complex combinations:
\be
\bos\equiv \f{1}{\sqrt{2}}({\bx}+i{\bp})\,,\;\;\;
\bosc\equiv \f{1}{\sqrt{2}}(\bx-i\bp)\,,\label{os}
\ee
\ni where we have settled $\hbar=1=m=\omega$ for simplicity. 
In the same manner, we will consider the complexified version $SU(1,1)$ of 
$SL(2,\Re)$ defined as 
\be
SU(1,1)\equiv\left\{ \hU= \left( \begin{array}{cc} \hz_1&\hz_2\\
\hz_2^*&\hz_1^*\end{array}
\right) ,\hz_i,\hz_i^* \in C/ \det(\hU)=|\hz_1|^2-|\hz_2|^2=1\right\} 
\ee 
\ni and the two-covering 
\be
SU(2)\equiv\left\{ U= \left( \begin{array}{cc} z_1&z_2\\-z_2^*&z_1^*\end{array}
\right) ,z_i,z_i^* \in C/ \det(U)=|z_1|^2+|z_2|^2=1\right\} 
\ee 
\ni of $SO(3)$ to account for half-integer values of the spin. 

Let us look at 
the structure of both groups as principal fibre bundles and choose a system 
of coordinates adapted to this fibration as follows:
\bea
&\heta &\equiv\f{\hz_1}{|\hz_1|}, \;\; \hc\equiv\f{\hz_2}{\hz_1},\;\;
\hcc\equiv\f{\hz_2^*}{\hz_1^*},\;\;\;\; 
\heta\in S^1,\;\;\hc,\hcc\in D_1\,;\nn \\
&\eta &\equiv\f{z_1}{|z_1|}, \;\; \c\equiv\f{z_2}{z_1},\;\;
\c^*\equiv\f{z_2^*}{z_1^*},\;\;\;\; \eta\in S^1,\;\;\c,\c^*\in S^2\,,\label{z}
\eea
\ni i.e., $SU(1,1)$ is a principal fibre bundle with fibre $U(1)$ and base 
the open unit disk $D_1$, whilst $SU(2)$ has the sphere $S^2$ as the base 
(to be precise, the coordinates $\c,\cc$, corresponding to a local chart 
at the identity, are related to the stereographical 
projection of the sphere on the plane). The action of $SU(1,1)$ on 
$\bos,\bosc$ can be written in matricial form as:
\be
\left(\begin{array}{c} \bos \\ \bosc \end{array}\right)\rightarrow 
\hU\left(\begin{array}{c} \bos\\ \bosc \end{array}\right)=
\sqrt{\f{1}{1-\hc\hc^*}}\left( \begin{array}{cc} 
\heta &\c\heta \\ \hcc\heta^{*}& \heta^{*}\end{array}\right)
\left(\begin{array}{c} \bos\\ \bosc \end{array}\right) 
\ee
\ni whereas the action of $SU(2)$ can be obtained by making use of the 
isometry between the $2\times 2$ hermitian matrices with  null trace 
$A\equiv\left(\begin{array}{cc} \os_3 & \os_1+i\os_2 \\ \os_1-i\os_2 & -\os_3
 \end{array}\right)$ and $\Re^3$ which leads to:
\bea
A&\rightarrow& UAU^\dag=\f{1}{1+\c\c^*}\left( \begin{array}{cc} 
\eta &\c\eta \\ -\cc\eta^{*}& \eta^{*}\end{array}\right)
\left(\begin{array}{cc} \os_3 & \os_1+i\os_2 \\ \os_1-i\os_2 & -\os_3
 \end{array}\right)\left( \begin{array}{cc} 
\eta^* &-\c\eta \\ \cc\eta^{*}& \eta\end{array}\right)\Rightarrow\nn\\
 \bos&\rightarrow& R\bos\,,
\eea
\ni where the correspondence $U\rightarrow R$ stands for the usual 
homomorphism between $SU(2)$ and $SO(3)$.

Now let us write, in compact form, the group law $\tg''=\tg'*\tg$ of 
the $13$-parameter Schr\"odinger quantizing group $\TG$ in $d=3$, 
which consists of a semidirect product 
$G={\bf C}^3\otimes_s \left( SU(2)\otimes SU(1,1)\right)$ 
suitably extended by $U(1)$ as follows:
\bea
\hU''&=&\hU'\hU\nn\\
\U''&=&\U'\U\;\;(\hbox{or}\;\; R''=R'R)\nn\\
\left(\begin{array}{c} \bos'' \\ {\bosc}'' \end{array}\right)&=&
\left(\begin{array}{c} \bos \\ \bosc \end{array}\right)+\hU^{-1}
\left(\begin{array}{c} R^{-1}\bos' \\ R^{-1}{\bosc}' \end{array}\right)
\label{leygrupo}\\
\zeta''&=&\zeta'\zeta\exp\f{1}{2}\left\{\left(\begin{array}{cc} \bos & 
\bosc \end{array}\right)\Omega\hU^{-1}\left(\begin{array}{c} R^{-1}\bos' 
\\ R^{-1}{\bosc}' \end{array}\right)\right\}\left(\eta''{\eta'}^{-1}
\eta^{-1}\right)^{2s}\left(\heta''{\heta'}{}^{-1}\heta^{-1}\right)^{2\hs}\,,\nn
\eea
\ni  where we denote $\Omega\equiv\left(\ba{cc} 0&1\\-1&0\ea\right)$ the 
central 
matrix in the Bargmann cocycle  and $s,\hs$ 
represent the spin and symplin indices related to both pseudo-extensions 
of $SU(2)$ and $SU(1,1)$ with generating functions 
$\delta(U)=s\theta,\,\theta\equiv -2i\log\eta\;\;$ 
and $\;\;\delta(\hU)=\hs\htheta,\,\htheta\equiv -2i\log\heta$, respectively. 
Note that both spin and symplin are forced to take half integer values 
\be
\s\equiv \f{k}{2},\;\;\;\hs\equiv\f{l}{2},\,\,k,l\in Z\label{halfint}
\ee
\ni only, for globality 
conditions (single-valuedness), as can be seen by expressing $\eta,\heta$ in 
terms of the global coordinates $z_i,\hz_i,\,i=1,2$ like in (\ref{z}) (see 
below).

The group law (\ref{leygrupo}) will be  the starting point for GAQ to obtain 
the irreducible representations of the Schr\"odinger group which, afterwards, 
we shall see,  correspond  with a 
three-dimensional isotropic harmonic oscillator carrying  two internal 
degrees of freedom. To this end, let us start writing the explicit 
expression of the left- and right-invariant vector fields:
\bea
\xl{\zeta}&=&\xr{\zeta}=\zeta\deriv{\zeta}\\
\xl{\bos} &=&\deriv{\bos}+\um\bosc\zeta\deriv{\zeta}\,,\;\;
\xl{\bosc} =\deriv{\bosc}-\um\bos\zeta\deriv{\zeta}\nn\\
\TXL_{\heta} &=&\heta\f{\p}{\p\heta}-2\hc\f{\p}{\p \hc}+2\hc^*
\f{\p}{\p \hc^*}-\bos\deriv{\bos}+\bosc\deriv{\bosc}\nn   \\
\TXL_{\hc} &=&-\f{1}{2}\heta \hc^* \f{\p}{\p\heta}+\f{\p}{\p \hc}-
\hcc{}^2\f{\p}{\p \hc^*}-\bosc\deriv{\bos}-\hs\hcc\zeta\deriv{\zeta}\nn \\
\TXL_{\hc^*} &=&\f{1}{2}\heta \hc \f{\p}{\p\heta}-\hc^2 \f{\p}{\p \hc}+
\f{\p}{\p \hc^*}-\bos\deriv{\bosc}+\hs\hc\zeta\deriv{\zeta}\nn \\
\TXL_{\eta} &=&\eta\f{\p}{\p\eta}-2\c\f{\p}{\p \c}+2\c^*
\f{\p}{\p \c^*}-2i(\os_2\deriv{\os_1}-\os_1\deriv{\os_2})-
2i(\osc_2\deriv{\osc_1}-\osc_1\deriv{\osc_2})\nn   \\
\TXL_{\c} &=&\f{1}{2}\eta \c^* \f{\p}{\p\eta}+\f{\p}{\p \c}+
{\c^*}^2\f{\p}{\p \c^*}+\s\cc\zeta\deriv{\zeta}\nn\\ &+&(\os_3\deriv{\os_1}-
\os_1\deriv{\os_3})+i(\os_2\deriv{\os_3}-\os_3\deriv{\os_2})+
(\osc_3\deriv{\osc_1}-\osc_1\deriv{\osc_3})+i(\osc_2\deriv{\osc_3}-
\osc_3\deriv{\osc_2})\nn \\
\TXL_{\c^*} &=&-\f{1}{2}\eta \c \f{\p}{\p\eta}+\c^2 \f{\p}{\p \c}+
\f{\p}{\p \c^*} -\s\c\zeta\deriv{\zeta}\nn\\ &+&(\os_3\deriv{\os_1}-
\os_1\deriv{\os_3})-i(\os_2\deriv{\os_3}-\os_3\deriv{\os_2})+
(\osc_3\deriv{\osc_1}-\osc_1\deriv{\osc_3})-i(\osc_2\deriv{\osc_3}-
\osc_3\deriv{\osc_2})\nn \\ 
& & \nn\\
\xr{\bos}&=&\hz^*_1R\deriv{\bos}-\hz^*_2R\deriv{\bosc}-\um(\hz^*_2R\bos+
\hz^*_1R\bosc)\zeta\deriv{\zeta}\nn\\
\xr{\bosc}&=&-\hz_2R\deriv{\bos}+\hz_1R\deriv{\bosc}+\um(\hz_1R\bos+
\hz_2R\bosc)\zeta\deriv{\zeta}\nn\\
\TXR_{\heta} &=&\heta \f{\p}{\p\heta} \nn \\
\TXR_{\hc} &=& \f{1}{2}\heta^{-1} \hc^* \f{\p}{\p\heta}+\heta^{-2}(1-\hc\hc^*)
\f{\p}{\p \hc}+\hs\heta^{-2}\hcc\zeta\deriv{\zeta}\nn \\
\TXR_{\hcc} &=&- \f{1}{2}\heta^3 \hc \f{\p}{\p\heta}+\heta^{2}(1-\hc\hc^*)
\f{\p}{\p\hcc}-\hs\heta^{2}\hc\zeta\deriv{\zeta}\nn\\
\TXR_\eta &=&\eta \f{\p}{\p\eta} \nn \\
\TXR_{\c} &=& -\f{1}{2}\eta^{-1} \c^* \f{\p}{\p\eta}+\eta^{-2}(1+\c\c^*)
\f{\p}{\p \c}-\s\eta^{-2}\cc\zeta\deriv{\zeta}\nn \\
\TXR_{\cc} &=& \f{1}{2}\eta^3 \c \f{\p}{\p\eta}+\eta^{2}(1+\c\c^*)
\f{\p}{\p\cc}+\s\eta^{2}\c\zeta\deriv{\zeta}\,.\nn
\eea
\ni The  (left) commutators between these vector fields are:
\be
\ba{l l l} 
\l[\xl{\zeta}, \hbox{all}\r]=0 &\l[\xl{\os_j},\xl{\osc_k}\r]=
-\delta_{jk}\xl{\zeta} & \l[\xl{\eta,\c,\cc},\xl{\heta,\hc,\hcc}\r]=0 \\ 
\l[\xl{\heta},\xl{\hc}\r]=2\xl{\hc}&\l[\xl{\heta},\xl{\hcc}\r]=
-2\xl{\hcc} & \l[\xl{\hc},\xl{\hcc}\r]=\xl{\heta}+2\hs\xl{\zeta} \\
\l[\xl{\eta},\xl{\c}\r]=2\xl{\c}&\l[\xl{\eta},\xl{\cc}\r]=
-2\xl{\cc} & \l[\xl{\c},\xl{\hc}\r]=-\xl{\eta}-2\hs\xl{\zeta} \\
\l[\xl{\heta},\xl{\bos}\r]=\xl{\bos}&\l[\xl{\hc},\xl{\bos}\r]=0 & 
\l[\xl{\hcc},\xl{\bos}\r]=\xl{\bosc} \\
\l[\xl{\heta},\xl{\bosc}\r]=-\xl{\bosc}&\l[\xl{\hc},\xl{\bosc}\r]=
\xl{\bos} & \l[\xl{\hcc},\xl{\bosc}\r]=0 \\
\l[\xl{\eta},\xl{\os_3}\r]=0 &\l[\xl{\eta},\xl{\os_2}\r]=
2i\xl{\os_1} & 
\l[\xl{\eta},\xl{\os_1}\r]=-2i\xl{\os_2} \\
\l[\xl{\c},\xl{\os_3}\r]=-\xl{\os_1}+i\xl{\os_2} &\l[\xl{\c},\xl{\os_2}\r]=
-i\xl{\os_3} & \l[\xl{\c},\xl{\os_1}\r]=\xl{\os_3} \\
\l[\xl{\cc},\xl{\os_3}\r]=-\xl{\os_1}-i\xl{\os_2} &\l[\xl{\cc},\xl{\os_2}\r]=
i\xl{\os_3} & \l[\xl{\cc},\xl{\os_1}\r]=\xl{\os_3} 
\ea\label{algebra}
\ee
\ni where we have omitted the commutators 
$\l[\xl{\eta,\c,\cc},\xl{\osc_j}\r]$, which have the same form 
as for the $\xl{\os_j}$ fields. The {\it Quantization 1-form} is:
\bea
\Theta &=& \f{i}{2}(\bosc d\bos -\bos d\bosc )\label{theta}\\ &+& 
\f{i}{2}\f{\hcc\bos\bosc +
{\bosc}^2}{1-\hc\hcc}d\hc -\f{i}{2}\f{\hc\bos\bosc +
{\bos}^2}{1-\hc\hcc}d\hcc +i\heta^*\f{(1+\hc\hcc)\bos\bosc +
\hc{\bosc}^2+\hcc{\bos}^2}{1-\hc\hcc}d\heta  \nn\\
 &-&i\f{\cc\L_3 -\L_-}{1+\c\cc}d\c +i\f{\c\L_3 -\L_+}{1+\c\cc}d\cc
+2i\eta^*\f{(1-\c\cc)\L_3 +\c\L_-+\cc\L_+}{1+\c\cc}d\eta\nn\\
 &+& \Theta_{SU(1,1)}+ \Theta_{SU(2)}-i\zeta^{-1}d\zeta\nn\\ 
\Theta_{SU(1,1)}&=&\f{i\hs}{1-\hc\hcc}(\hc d\hcc -\hcc d\hc -
4\hc\hcc\heta^* d\heta)\nn\\
\Theta_{SU(2)}&=&\f{i\s}{1+\c\cc}(-\c d\cc+\cc d\c +
4\c\cc\eta^* d\eta)\,,\nn
\eea
\ni where we have denoted ${\bf \L}\equiv i(\bosc\times\bos)$ and 
$\L_\pm\equiv L_1\pm i\L_2$. The {\it characteristic module} 
$\hbox{Ker}\Theta\cap \hbox{Ker} {}d\Theta$ is generated by the left 
subalgebra:
\be
{\cal G}_\Theta=<\xl{\heta},\,\,\xl{\eta} >\,,\label{char}
\ee 
\ni the trajectories of which represent the generalized Hamiltonian equations 
of motion on the $10$-dimensional symplectic manifold 
${\bf C}^3\otimes D_1\otimes S^2$ of the theory. The Noether invariants under 
these equations are:
\be
\ba{ll}
F_{\bos}\equiv i_{\TXR_{\bos}}\Theta=i(\hz^*_1R\bosc +\hz^*_2R\bos)& 
F_{\bosc}\equiv i_{\TXR_{\bosc}}\Theta=-i(\hz_1R\bos +\hz_2R\bosc)\\
F_{\heta}\equiv i_{\TXR_{\heta}}\Theta=\hs\f{-4i\hc\hcc}{1-\hc\hcc}+i
F_{\bosc}\cdot F_{\bos} & F_{\eta}\equiv i_{\TXR_{\eta}}\Theta=
\s\f{4i\c\cc}{1+\c\cc}+2(F_{\bosc}\times F_{\bos})_3\\
F_{\hc}\equiv i_{\TXR_{\hc}}\Theta=\hs\f{-2i\heta^{-2}\hcc}{1-\hc\hcc}-
\f{i}{2}F_{\bos}\cdot F_{\bos} &F_{\c}\equiv i_{\TXR_{\c}}\Theta=
\s\f{2i\eta^{-2}\cc}{1+\c\cc}+(F_{\bosc}\times F_{\bos})_-\\
F_{\hcc}\equiv i_{\TXR_{\hcc}}\Theta=\hs\f{2i\heta^{2}\hc}{1-\hc\hcc}+
\f{i}{2}F_{\bosc}\cdot F_{\bosc} &
F_{\cc}\equiv i_{\TXR_{\cc}}\Theta=\s\f{-2i\eta^{2}\c}{1+\c\cc}+
(F_{\bosc}\times F_{\bos})_+\,.\label{noether}
\ea
\ee
\ni These Noether invariants parametrize the classical manifold of the 
corresponding quantum system [note that the non-dynamical (non-basic) Noether 
invariants $F_{\eta}$ and $F_{\heta}$, coming from the vector fields the 
left version of which are in the 
characteristic subalgebra (\ref{char}), are expressed in terms of the  
rest (the basic ones)]. One can naturally define a Poisson braket as 
\be
\left\{ F_{\tg^j},F_{\tg^k}\right\}
\equiv i_{\l[ \xr{\tg^j},\xr{\tg^k}\r] }\Theta\label{poisson1}
\ee
\ni which, according to the Lie algebra (\ref{algebra}), reproduces the 
standard expressions in terms of 
\be
\{1,\,\bos,\,\bosc,\,-\f{i}{2}\bos^2,\,\f{i}{2}{\bosc}^2,\,i\bosc\bos,\, 
\bosc\times\bos\}\label{poisson2}
\ee
\ni for $\hs=\s=0$ only. Also, for these particular (classical) values, all 
Noether invariants are expressed in terms of the basics $F_{\bos}$ and 
$F_{\bosc}$ (as it can be easily seen in (\ref{noether})), obtaining a 
new reduction of the symplectic manifold 
\be
{\bf C}^3\otimes D_1\otimes S^2\rightarrow {\bf C}^3\label{reduction}
\ee
\ni from $10$ to $6$ dimensions (i.e., losing internal 
degrees of freedom). From the quantum point of view, this reduction is 
due to the enlarging 
of the characteristic subalgebra (\ref{char}) which, now, incorporates the 
whole $su(2)$ and $su(1,1)$ subalgebras.

Until now, the way to address both groups $SU(2)$ and $SU(1,1)$ has been 
rather parallel. The difference starts when we look for a full-polarization 
subalgebra (\ref{defpol}) intended to reduce the representation 
(\ref{tcondition}) for 
the $\s=\hs=0$ case, i.e., when we try to represent irreducibly and unitarily 
the classical Poisson algebra (\ref{poisson2}) on 
a Hilbert space of wave functions depending arbitrarily on half of the 
symplectic variables, let us say, 
the $\bosc$ coordinates only. As can be easily checked in (\ref{algebra}), 
whereas the $su(2)$ subalgebra of ${\cal G}_\Theta$ is diagonal under 
commutation 
with either $\xl{\bos}$ or $\xl{\bosc}$ (closing a horizontal subalgebra 
separately), the $su(1,1)$ subalgebra is not; i.e., it mixes 
$\xl{\bos}$ and $\xl{\bosc}$ and precludes a full-polarization subalgebra 
for this case. This obstruction is a particular example of what we have 
already defined as an algebraic anomaly and shares with the more conventional 
characterization the appearance of central charges in the quantum algebra of 
operators. The standard quantization solves this problem by imposing normal 
order by hand, leading to a quantum algebra differing from the classical 
(\ref{poisson2}) Poisson algebra by central (normal order) 
terms and providing an irreducible 
representation of the metaplectic group $Mp(2,\Re)$ (two-cover of 
$Sp(2,\Re)\simeq SU(1,1)$) with Bargmann index $k=3/4$ ($=d/4$ in $d$ 
dimensions).
This situation can be seen as a ``weak" 
(avoidable) violation of the {\it no-go} theorems, and we shall show 
in Sec. IV that one can, in fact, go further.  

Let us show how GAQ solves this obstruction (reduction of the quantum 
representation) by means of higher-order 
polarizations, the existence of which will be guaranteed only for the 
particular 
(quantum) value of $\hs=-\f{3}{4}$ [$\hs=-\f{d}{4}$ in $d$ dimensions], as 
opposed to the classical value of $\hs=0$ (for which the counterpart 
classical reduction (\ref{reduction}) is achieved). To this end, let us 
firstly 
calculate the irreducible representations of the Schr\"odinger group 
with arbitrary spin and symplin and then show how the aforementioned reduction 
takes place. 

A full-polarization subalgebra exists for arbitrary (non-zero) $\s$ and $\hs$ 
which is:
\be
{\cal P}=<\xl{\eta},\,\xl{\heta},\,\xl{\c},\,\xl{\hc},\,\xl{\bos}>\,.
\label{pola}
\ee
\ni The general solution to the polarization equations (\ref{defpol}) leads to 
a Hilbert space ${\cal H}^{(\s,\hs)}(\TG)$ of wave functions of the form:
\bea
\psi^{(\s,\hs)}(\zeta,\eta,\c,\cc,\heta,\hc,\hcc,\bos,\bosc)&=&\zeta(1+
\c\cc)^{-\s}
(1-\hc\hcc)^{-\hs}e^{-\um (\bosc\bos +\hc{\bosc}^2)}\phi(\chi,\hchi,\bb)\nn\\
\chi&\equiv&\eta^{-2}\cc,\;\;\;\hchi\equiv\heta^{-2}\hcc,\;\;\;
\bb\equiv (1-\hc\hcc)^\um\heta^*R\bosc\,.\label{wavepol}
\eea
\ni A scalar product can be given through the invariant integration volume 
(\ref{volumen}) of $\TG$:
\bea
v(\tg)&=&i\f{1}{(1+\c\cc)}\f{1}{(1+\hc\hcc)}
\l[\prod_{j=1}^3d\hbox{Re}(\os_j)\wedge d\hbox{Im}(\os_j)\r]\nn\\
&\wedge& \l[ d\hbox{Re}(\c)\wedge d\hbox{Im}(\c)\wedge \eta^{-1}d\eta\r]\wedge
\l[ d\hbox{Re}(\hc)\wedge d\hbox{Im}(\hc)\wedge \heta^{-1}d\heta\r]\wedge
 \zeta^{-1}d\zeta\,.\label{volucho}
\eea
\ni Let us call 
\be
\psi^{(m,\hm)}_{\bn}\equiv \zeta(1+\c\cc)^{-\s}
(1-\hc\hcc)^{-\hs}e^{-\um (\bosc\bos +\hc{\bosc}^2)}(\chi)^{m}(\hchi)^{\hm}
(\b_1)^{n_1}(\b_2)^{n_2}(\b_3)^{n_3}\label{basic} 
\ee
\ni a basic wave function where $m$ and $\hm$ stand for the third components 
of spin and symplin, respectively, and $n_j$ represents the oscillator quanta 
in the $j$ direction. The requirement of analyticity of these basic wave 
functions when expressed in terms of global coordinates $z_i,\hz_i,\,i=1,2$ 
(as in 
(\ref{z})) leads to integrality conditions  $2\s,2\hs,m,\hm\in Z$, where we 
recover the conditions in (\ref{halfint}). 
The action of the right-invariant vector fields 
(operators) on ${\cal H}^{(\s,\hs)}(\TG)$ can be given
 on these basic functions 
as follows (we write $\hej\equiv 
(\delta_{1,j},\delta_{2,j},\delta_{3,j})$):
\bea
\xr{\os_j}\psi^{(m,\hm)}_{\bn}&=&-\psi^{(m,\hm)}_{\bn+\hej}-
n_j\psi^{(m,\hm+1)}_{\bn-\hej}\label{repre}\\
\xr{\osc_j}\psi^{(m,\hm)}_{\bn}&=& n_j\psi^{(m,\hm)}_{\bn-\hej}\nn\\
\xr{\heta}\psi^{(m,\hm)}_{\bn}&=&-(2\hm+\sum_j n_j)\psi^{(m,\hm)}_{\bn}\nn\\
\xr{\hc}\psi^{(m,\hm)}_{\bn}&=&-(\hm-2\hs+\sum_j n_j)
\psi^{(m,\hm+1)}_{\bn}-\um\sum_j\psi^{(m,\hm)}_{\bn+2\hej}\nn\\
\xr{\hcc}\psi^{(m,\hm)}_{\bn}&=&\hm\psi^{(m,\hm-1)}_{\bn}\nn\\
\xr{\eta}\psi^{(m,\hm)}_{\bn}&=&-2m\psi^{(m,\hm)}_{\bn}+2i(
n_1\psi^{(m,\hm)}_{\bn+\hed-\heu}-n_2\psi^{(m,\hm)}_{\bn+\heu-\hed})\nn\\
\xr{\c}\psi^{(m,\hm)}_{\bn}&=&-(2\s-m)\psi^{(m+1,\hm)}_{\bn}-(
n_1\psi^{(m,\hm)}_{\bn+\het-\heu}-n_3\psi^{(m,\hm)}_{\bn+\heu-\het}) \nn \\
 & & -i(n_3\psi^{(m,\hm)}_{\bn+\hed-\het}-n_2\psi^{(m,\hm)}_{\bn+\het-\hed})
\nn\\
\xr{\cc}\psi^{(m,\hm)}_{\bn}&=&m\psi^{(m-1,\hm)}_{\bn}-(
n_1\psi^{(m,\hm)}_{\bn+\het-\heu}-n_3\psi^{(m,\hm)}_{\bn+\heu-\het})
+i(n_3\psi^{(m,\hm)}_{\bn+\hed-\het}-n_2\psi^{(m,\hm)}_{\bn+\het-\hed})\,.\nn
\eea
\ni From these expressions we conclude that the third component of spin 
$m$ is restricted to take 
the values $m=0,...,2\s$ (finite dimensional subspace), whilst $\hm$ can 
take any value from $0$ to $\infty$ (infinite dimensional subspace), the 
difference being the compact  and non-compact character of the corresponding 
subgroups $SU(2)$ and $SU(1,1)$. Thus, any wave function in 
${\cal H}^{(\s,\hs)}(\TG)$ can be expressed as an arbitrary 
linear combination of these basic wave functions:
\be
\psi^{(\s,\hs)}=\sum^3_{j=1}\sum^\infty_{n_j=0}\sum^\infty_{\hm=
0}\sum^{2\s}_{m=0}
c^{(m,\hm)}_{\bn}\psi^{(m,\hm)}_{\bn}\,.
\ee
\ni Note that the set of wave functions $\{\psi^{(m,\hm)}_{\bn}\}$ is not 
orthogonal under the scalar product (\ref{volucho},\ref{scalarp}), but can be 
expressed in terms of an orthogonal set $\{\check{\psi}^{(m,\hm)}_{\bk}\}$ 
as follows:
\bea
\psi^{(m,\hm)}_{\bn}&=&\sum^{\l[\f{n_1}{2}\r]}_{l_1=0}
\sum^{\l[\f{n_2}{2}\r]}_{l_2=0}\sum^{\l[\f{n_3}{2}\r]}_{l_3=0}
\left(\prod^3_{j=1}(-1)^{n_j+l_j}\Gamma_{n_j,l_j}\right)
\check{\psi}^{(m,\hm+l_1+l_2+l_3)}_{\bn-2{\bf l}}\\
\Gamma_{n_j,l_j}&\equiv&\f{1}{2^{l_j}}\f{(n_j+2l_j)!}{n_j!l_j!}\nn\\
\check{\psi}^{(m,\hm)}_{\bk}&\equiv&\left(\xr{\os_1}\right)^{k_1}
\left(\xr{\os_2}\right)^{k_2}\left(\xr{\os_3}\right)^{k_3}
\psi^{(m,\hm)}_{{\bf 0}}\,,\nn
\eea
\ni where $\l[\f{n_j}{2}\r]$ stands for the integer part of $\f{n_j}{2}$.

Let us define the {\it intrinsic} (internal) higher-order operators:
\be
\ba{ll} {\xr{\heta}}{}^{(HO)}\equiv \xr{\heta}-\xr{\bos}\cdot\xr{\bosc} & 
{\xr{\eta}}{}^{(HO)}\equiv \xr{\eta}+2i\left(\xr{\bos}\times
\xr{\bosc}\right)_3\\
{\xr{\hc}}{}^{(HO)}\equiv \xr{\hc}+\um\xr{\bos}\cdot\xr{\bos} &
{\xr{\c}}{}^{(HO)}\equiv \xr{\c}+i\left(\xr{\bos}\times\xr{\bosc}\right)_-\\
{\xr{\hcc}}{}^{(HO)}\equiv \xr{\hcc}-\um\xr{\bosc}\cdot\xr{\bosc} &
{\xr{\cc}}{}^{(HO)}\equiv \xr{\cc}-i\left(\xr{\bos}\times\xr{\bosc}\right)_+
\ea \label{hovf}
\ee
\ni which close a Lie subalgebra of the right enveloping algebra of the 
Schr\"odinger group $\TG$, isomorphic to the Lie algebra 
$su(2)\oplus su(1,1)$ with a particular pseudo-extension:
\be
\l[{\xr{\c}}{}^{(HO)},{\xr{\cc}}{}^{(HO)}\r]={\xr{\eta}}{}^{(HO)}+2\s\,,\;\;\;
\l[{\xr{\hc}}{}^{(HO)},{\xr{\hcc}}{}^{(HO)}\r]=-{\xr{\heta}}{}^{(HO)}-
(2\hs+\f{3}{2})\,.\label{ideal}
\ee
\ni They represent the observables corresponding to the (pure) 
internal degrees of freedom: symplin and spin. Even more, 
this subalgebra proves to be, in general, an ideal (under commutation) of 
the right enveloping algebra $U\tilde{{\cal G}}^R$ of $\TG$, and a 
horizontal ideal for the particular values  
\be
\hs=-\f{3}{4}, \,\;\;\hbox{and} \;\;\; \s=0 \label{ano}
\ee
\ni (as can be partially checked in (\ref{ideal})). This last situation 
requires special attention. In fact, the existence of a non-trivial 
(non zero) horizontal ideal --gauge subalgebra 
(see the paragraph after (\ref{camposlr}))-- 
is a sign of reducibility; indeed, according to general settings 
\cite{config2}, the right-invariant vector fields in a 
gauge subalgebra must be a linear combination of left-invariant vector 
fields in the characteristic module and, therefore, they have to be trivially 
represented (zero). Then, the representation 
(\ref{repre}) is reducible for the particular {\it quantum} values 
(\ref{ano}), as opposed to the {\it classical} values 
(concerning the symplin but not the spin) for which the classical reduction 
(\ref{reduction}) was achieved.  Nevertheless, whereas the (normal) 
reduction for $\s=0$ is reached simply by a new full 
polarization consisting of (\ref{pola}) enlarged by $\xl{\cc}$ 
(i.e. containing the whole $su(2)$ subalgebra of the characteristic algebra), 
the (anomalous) reduction for $\hs=-3/4$ requires to use higher-order 
polarization techniques. The {\it modus operandi} to 
construct  a higher-order polarization subalgebra for these anomalous 
cases usually consists in  deforming the generators in the 
characteristic subalgebra 
corresponding to the classical reduction of the symplectic manifold 
($su(1,1)$ in our case), 
by adding terms  in the  left enveloping algebra $U\tilde{{\cal G}}^L$. 
Also, when there are non-trivial higher-order gauge operators, their 
(equivalent) 
left counterparts are candidates for further reducing the representation. 
In our case, a new higher-order restriction on wave functions (\ref{wavepol}) 
can be consistently added to the set of first-order restrictions given by 
(\ref{pola}) for the anomalous case $\hs=-3/4$ only.  
The candidate for this reduction process is the deformation 
${\xl{\hcc}}{}^{(HO)}$ of  $\xl{\hcc}$, which is precisely 
the left counterpart of ${\xr{\hcc}}{}^{(HO)}$ defined in (\ref{hovf}). Its 
gauge character  makes indifferent if the higher-order polarization condition 
\be
{\xl{\hcc}}{}^{(HO)}\psi^{(\s,-\f{3}{4})}=0 \;\;\Rightarrow 
\f{\p\phi}{\p\hchi}-\um\sum_j\f{\p^2\phi}{\p\b^2_j}=0\label{horest}
\ee
\ni is imposed as a left or as a right restriction 
on wave functions (\ref{wavepol}), the solution of which is expressed in 
terms of an 
orthogonal and complete set of the form:
\be
\check{\psi}^{(m)}_{\bk}\equiv \sum_{i=1}^3\sum_{l_i=0}^\infty 
\sum_{n_i=0}^\infty\left(\prod^3_{j=1}\Gamma_{n_j,l_j}\delta_{n_j+2l_j,k_j}
\right)\psi^{(m,l_1+l_2+l_3)}_{\bn}=\left(\prod^3_{j=1}(-1)^{k_j}
\left(\xr{\os_j}\right)^{k_j}\right)\psi^{(m,0)}_{{\bf 0}}\,,\label{hermite}
\ee
i.e., the orbit  of the creation operators 
$\ac_j\equiv -\xr{\os_j}$ through the vacuum 
$\psi^{(0,0)}_{{\bf 0}}$ (when $\s=0$). 
In this way, the whole set of physical operators 
$\xr{\tg^i}$ are expressed in terms of basic ones $\ac_j=-\xr{\os_j}$ and 
$\a_j\equiv\xr{\osc_j}$ as in (\ref{hovf}) taking into account that now 
(internal) 
higher-order operators ${\xr{\tg^i}}{}^{(HO)}$ are trivially zero (gauge). 
For example, the energy operator is:
\be
\hat{E}\equiv -X^R_{\heta}\equiv -\xr{\heta}-2\hs\xr{\zeta}=
        \sum_{j=1}^3(\ac_j\a_j+\um)\;,
\ee
\ni where the last redefinition of the $\xr{\heta}$ generator is intended 
to render
the commutation relation $[\xr{\hc},\xr{\hcc}]=\xr{\heta}+2\hs\xr{\zeta}$ in
(\ref{algebra}) to the usual $su(1,1)$ one: 
$[X^R_{\hc},X^R_{\hcc}]=X^R_{\heta}$.
Note that the zero-point energy $E_0=\um d$ of the harmonic 
oscillator is precisely 
$-2\hs$ for the quantum value of the anomaly $\hs=-\frac{d}{4}$. 
This anomalous value
is obtained in the standard approach by the na\"\i ve ``symmetrization rule" 
and proves to have important physical consequences in the 
experimentally observed {\it Casimir effect} (see \cite{Casimir} 
and references 
therein).

The half-half integer character of the symplin $\hs=-\um(\f{3}{2})$ indicates, 
according to (\ref{halfint}), that the representation of $SU(1,1)$ is 
{\it bi-valuated}; i.e. it is the two-cover $Mp(2,\Re)$ (metaplectic group) 
which is in fact faithfully represented. 

At this stage, a comparison of the fundamentals  of  this finite-dimensional 
anomalous system with the more conventional one 
(infinite-dimensional) of the bosonic string \cite{cuerda} is opportune. 
The role played by the Virasoro group, acting on string modes 
$\{\alpha^\mu_m\}$,
\be 
\l[\hat{L}_n,\hat{\alpha}^\mu_m\r]=m\hat{\alpha}^\mu_{n+m}
\ee
\ni is here played by $SU(1,1)$ acting on  
oscillator modes $\bos,\bosc$. Like the $su(1,1)$ algebra, the Virasoro 
algebra 
\be
\l[\hat{L}_n,\hat{L}_m\r] =(n-m)\hat{L}_{n+m}+
\f{1}{12}(cn^3-c'n)\delta_{n,-m}\hat{1}
\ee
\ni appears also centrally extended, although, this time, 
by both a pseudo-cocycle (with parameter $c'$ generalizing the symplin 
$\hs$) 
and a true cocycle (with parameter $c$); the latter is permitted 
by the infinite-dimensional character of the Virasoro group, 
which violate the Whitehead Lemma \cite{cabezablanca}. The 
(anomalous) reduction which allows the Virasoro operators to 
be written in terms of the string modes (Sugawara's construction 
\cite{Sugawara})
 for $c=c'=d$ (the dimension of the space-time) 
\be
\hat{L}_k=\um g_{\mu\nu}:\sum\hat{\alpha}^\mu_{k-n}\hat{\alpha}^\nu_n :
\label{Sugawara}
\ee
\ni is essentially equivalent (in nature) 
to the anomalous reduction which allows the $su(1,1)$ 
operators to be written in terms of the oscillator modes 
for $\hs=-d/4$. In fact, formula (\ref{hermite}) expressing the states
of the invariant, reduced subspace of the Hilbert space 
${\cal H}^{(\s,\hs=-\f{3}{4})}(\TG)$ as generated  by 
the action of the creation operators corresponding 
to the harmonic oscillator only, parallels the construction of the reduced 
Hilbert space of the string by the action of just string mode operators on 
the vacuum (see e.g. \cite{cuerda})
\be
\hat{\alpha}^{\mu_1}_{n_1}\hat{\alpha}^{\mu_2}_{n_2}...
\hat{\alpha}^{\mu_i}_{n_i} |0\rangle\,.
\ee

To proceed further in this comparison, 
we could simulate the constraints in string theory by  
restricting our  oscillator to the sphere; more precisely, the role played 
by the 
Virasoro group generators, acting as constraints in string theory, 
can be played here by part of the $su(1,1)$ operators, for example, 
those which restrict 
the particle to move on the sphere. Indeed, making use of the expression 
(\ref{os}) we can write the square of the vector position operator and 
its ``time derivative'' as: 
\bea
\hat{\bx}^2&\equiv & \um ((\xr{\bosc})^2+(\xr{\bos})^2-\xr{\bosc}\xr{\bos}
-\xr{\bos}\xr{\bosc})=\xr{\hcc}-\xr{\hc}-\xr{\heta}+\f{3}{2}\\
\um (\hat{\bx}\hat{\bp}+\hat{\bp}\hat{\bx})&\equiv &-
\f{i}{2}((\xr{\bosc})^2-(\xr{\bos})^2)=-i(\xr{\hcc}+\xr{\hc})\,.\nn
\eea
\ni  The constrained theory can be formulated by looking at the Schr\"odinger 
group as a principal fibre bundle where the structure group $T=U(1)$ has been 
replaced by $T=\tilde{A}(1)$, a central extension of the Affine 
group in 1D; more precisely, the Lie algebra of $T$ is now:
\bea
\l[\xr{\tilde{t}_1},\xr{\tilde{t}_2}\r] =2i\xr{\tilde{t}_1}+
2ir^2\xr{\zeta}\label{liga}\\
\xr{\tilde{t}_1}\equiv \hat{\bx}^2-r^2\xr{\zeta}\,,\;\;\;
\xr{\tilde{t}_2} \equiv \um (\hat{\bx}\hat{\bp}+\hat{\bp}\hat{\bx})\,,\nn
\eea
\ni which takes part of a subalgebra of $su(1,1)$ pseudo-extended by $U(1)$ 
with parameter $r$ (radius of the sphere). The constraint on the sphere 
can be achieved through $T$-equivariant conditions (\ref{const}) 
on arbitrary combinations $\check{\psi}^{(s)}$ of the basic wave functions 
(\ref{hermite}), either as
\be 
 \xr{\tilde{t}_1}\check{\psi}^{(s)}=0\;\;\;
\hbox{or}\;\;\; \xr{\tilde{t}_2}\check{\psi}^{(s)}=0\,,\label{tequiv}
\ee
\ni since the conjugate character of these two constraints (see the commutator 
in (\ref{liga})) prevents fixing both at a time, i.e. $T_p$ is generated 
by either $\xr{\tilde{t}_1}$ or $\xr{\tilde{t}_2}$. 

With regard to the good operators of the theory, there
are some sets of operators which preserve one option of 
$T_p\subset T_B\subset T$, but not the other. We shall restrict ourselves
to the intersection of these sets to define our good operators, i.e. those
operators which preserve any of the possible choices of $T_B\subset T$. This
set of good operators is enough to reproduce the constrained quantum system
of the particle on the sphere; it is:
\bea
\tilde{{\cal G}}_T&=& \left\{\xr{\bos}\times\xr{\bosc},\,\xr{\eta},\,\xr{\c},
\,\xr{\cc},\, \hat{u}_1,\,
\hat{u}_2,\,\hat{u}_3\right\}\\
\hat{{\bf u}}&\equiv& \hat{r}^{-1}\hat{\bx}\,,\;\;\;
 \hat{r}^{-1}\equiv \f{1}{r}\left(1+\sum^\infty_{n=1}(-1)^n
\f{(2n-1)!!}{2n!!}\left(\f{\xr{\tilde{t}_1}}{r^2}\right)^n\right)=
\f{1}{\sqrt{r^2+\xr{\tilde{t}_1}}}
\nn
\eea
\ni which commute with both $\xr{\tilde{t}_1}$ and $\xr{\tilde{t}_2}$ and 
close a subalgebra isomorphic to the Euclidean algebra in 3D for the case 
of spin $s=0$ 
[note that the operators $\hat{u}_j$ live 
in the enveloping algebra of the Schr\"odinger group]. When we choose the 
second option in (\ref{tequiv}), the constrained  
Hilbert space  turns out to be made up of $T_B$-equivariant functions 
constructed by taking the orbit of $\hat{u}_j$ through the only $2\s+1$ 
states that are  ``rotationally invariant'' 
and annihilated by $\xr{\tilde{t}_2}$. These prove to be:
\be
Y^{(0)}_{m_s,0}\equiv \sum^\infty_{q=0}\f{K}{(2q)!!(2q-N_0)!!}
\left(\xr{\c}\right)^{2q}\check{\psi}^{(m_s)}_{{\bf 0}}\,,\;\;\;
m_s=0,...,2s+1
\ee
\ni where $K$ is an arbitrary constant and $N_0=2\hs+1=-\um$. 
For $s=0$, the state $Y^{(0)}_{0,0}$ simply represents the spherical 
harmonic of  zero angular momentum [note that this state is an infinite 
linear combination of harmonic oscillator wave functions]. States with 
higher values of angular momentum $Y^{(l)}_{m_s,m_l}$ correspond to 
the repeated action of $\hat{u}_j$ on these ``vacua'' 
$Y^{(0)}_{m_s,0}$. For example,  the state 
$Y^{(1)}_{m_s,0}\equiv\hat{u}_3Y^{(0)}_{m_s,0}$ has (orbital) angular 
momentum $l=1$ and third component $m_l=0$. The whole set of states obtained 
in this way represent the Hilbert space of a spinning point particle 
living on the sphere. Different values of $\s$ parametrize    
non-equivalent quantizations. 

We have preferred to maintain the internal degree of spin in order to 
make comparisons with other  approaches to {\it Quantum mechanics on $S^D$} as 
Ref. \cite{ohnuki}, where the $(D+1)$-dimensional Euclidean group was  
used to study the point particle on $S^D$, or Refs. 
\cite{landsman,macmullan}, where $S^D$ is seen as a coset space 
$G/H=SO(D+1)/SO(D)$ of $SO(D+1)$. An important basic 
difference of our procedure with respect to other approaches is that the 
sphere 
$S^2$ where the ``free particle" of \cite{ohnuki,landsman,macmullan} lives,  
seems to correspond with our {\it internal} sphere 
$S^2$ immersed on the symplectic manifold  $T^*S^2\times S^2$ [$T^*S^2$ is the 
cotangent of $S^2$], resulting from the original 
${\bf C}^3\otimes D_1\otimes S^2$ (see (\ref{reduction})) after reducing 
(\ref{horest}) [$\rightarrow {\bf C}^3\otimes S^2$] and constraining 
(\ref{tequiv});  i.e., there are {\it two} different (in nature) 
spheres for us, a ``real" sphere 
immersed in $\Re^3$, where the particle lives, and an ``internal" (symplectic) 
sphere $S^2=SU(2)/U(1)$ corresponding to the  spin degree 
of freedom. This situation 
can lead to confusions in interpretation when they quantize 
on coset spaces $Q=G/H$ and  parametrize $Q$ as immersed in $\Re^n$;  
in fact, a embedding of our $Q=S^2=SU(2)/U(1)$ in $\Re^3=\{y_1,y_2,y_3\}$ 
according to a standard stereographical projection map:
\be
\c\eta^2=\f{y_1}{\rho+y_3}+i\f{y_2}{\rho+y_3}\,,\;\;\;\hbox{with}\;\;\;
{\bf y}^2=\rho^2
\ee
\ni could lead us to believe that ``a monopole is present" if we 
interpret the 1-form connection $\Theta_{SU(2)}$ in (\ref{theta}) as 
a $U(1)$-gauge potential [it is called the {\it $H$-connection} or 
{\it cannonical connection} in 
\cite{landsman,macmullan}], but we know that ``this monopole does not  live 
in our world".

\section{Breaking through no-go theorems: extended objects from the 
elementary particle}

As already mentioned, the Schr\"odinger algebra can be viewed as the maximal
Poisson subalgebra on the solution manifold of the free particle and/or the 
harmonic oscillator that can be
quantized in a more or less canonical way. This means that the 
quantization mapping ``$\;\hat{}\;$'' representing the Poisson subalgebra 
$<1,x,p,\um x^2,\um p^2,xp>$ by $<\hat{1},\hat{x},\hat{p},\um \hat{x}^2,
\um \hat{p}^2,\hat{x}\hat{p}>$ is not a Lie algebra homomorphism due to the 
(anomalous) term $-\frac{i}{2}$ in the commutator 
$[\um \hat{x}^2,\um \hat{p}^2]$, with regard to its associated 
Poisson bracket. Fortunately, this anomaly can easily
be hidden simply by symmetrizing the operator $\widehat{(xp)}$.

Standard canonical quantization fails to go beyond any Poisson 
subalgebra containing polynomials in $x,p$ of degree greater than two 
\cite{Folland,no-go}.
From the point of view of group quantization, however, we can proceed
further, provided that we are able to close a definite Poisson 
subalgebra that, although necessarily infinite-dimensional, has a controlled 
growing (finite growth; see for instance \cite{Serre}). Then a group 
law can be found, at least, by exponentiating the Lie algebra 
order by order, as in the case, for instance, of Kac-Moody algebras 
\cite{Formal}, and by considering all possible (pseudo-)extensions (and 
associated deformations) with arbitrary parameters $\gamma_k$. 

Needless to say, in the quantization process many
anomalies will eventually appear, requiring the use of the higher-order
polarization technique. These anomalies are really obstructions to the 
quantization of given functions of $x,p$ in terms of $\hat{x},\hat{p}$.
The quantum values $\gamma_k^{(0)}$ of the anomalies are precisely those 
for which such 
a task can be achieved even though the 
quantization morphism ``$\;\hat{}\;$'' is somewhat
distorted (central terms for operators representing
quadratic functions and more general terms for higher-order 
polynomials on $x,p$). Far from the quantum values of the anomalies, 
however, new (purely) quantum degrees of freedom must enter the theory 
as associated with those operators which cannot be expressed in terms 
of $\hat{x},\hat{p}$. Moreover, it could well happen that no quantum values  
of $\gamma_k$ exist for some cases, thus leading to ``essentially anomalous" 
(inescapable) situations. 

To construct such an infinite-dimensional Poisson algebra, 
generalizing the Schr\"odinger algebra, 
let us start with the solution manifold of the elementary particle in two  
dimensions parametrized by $x,p$. For simplicity, we shall asume that
the particle is non-relativistic, although we could think of the 
relativistic situation so long as $x$ really represents the classical
analogue of the Newton-Wigner position operator \cite{N-W,Position}; or, 
we could even also consider the time parameter $x^0$, provided that it is
given a dynamical character with canonically conjugate momentum $p^0$ 
($\{x,p\}=1\rightarrow\{x^\mu,p^\nu \}=g^{\mu\nu}$), and
then impose the mass-shell constraint \cite{VAL}. 
Let us continue to use an oscillator-like parametrization of the phase 
space, as in (\ref{os}), and choose the following set of classical functions 
of $\osc,\os$:

\bea
\A_n^\alpha = \um \os^{2n}({\osc}\os)^{-\alpha-n+1}\,,\;\;\;
\A_{-m}^\beta = \um {\osc}^{2m}({\osc}\os)^{-\beta-m+1}\label{motivation}\\
n,m=0,\um,1,\f{3}{2},...
\;\;\;\alpha,\beta=0,\pm\um,
\pm 1,\pm\f{3}{2},...\nn
\eea
\ni which generalize the Virasoro algebra (as generating diffeomorphisms of 
the plane) and contain the Schr\"odinger algebra as the largest 
finite-dimensional 
subalgebra.  
 
A straightforward computation from the basic Poisson bracket $\{{\osc},\os\}=i$
provides the following formal, Poisson algebra:
\be
\{\A_n^\alpha,\A_m^\beta\}=-i[(1-\beta)n-(1-\alpha)m]\A_{n+m}^{\alpha+\beta}
      \;\;\;n,m,\alpha,\beta\in Z/2 \label{Auaral}
\ee
\ni which should not be confused with that introduced in \cite{Floratos}.
It contains some interesting subalgebras: 

\ni {\it Schr\"odinger algebra}:
\be
\ba{llllll} {\osc}=2\A_{-\um}^{\um}, & \os=2\A_{\um}^{\um}, & 1\equiv 
2\A_0^1, &
\um {\osc}^2=\A_{-1}^0, & \um \os^2=\A_1^0, & {\osc}\os=2\A_0^0\,.
\ea
\ee
\ni {\it Virasoro algebra}:
\be 
L_n\equiv\A_n^0\Rightarrow \left\{L_n,L_m\right\}=-i(n-m)L_{n+m}
\ee
\ni {\it Unextended ``string" algebra}: The already identified Virasoro 
subalgebra can be enlarged by
$\alpha_m\equiv L_m^1,\;m\in Z$. They close the following semi-direct
algebra:
\bea
\{L_n,L_m\}&=&-i(n-m)L_{n+m} \label{Guita} \label{string}\\
\{L_n,\alpha_m\}&=&im\alpha_{n+m}\nn\\
\{\alpha_n,\alpha_m\}&=&0 \,,\nn
\eea
\ni corresponding to the (classical) underlying symmetry  of 
string theory (one for each value of the the $\mu$ index
in $\alpha_m^\mu$), i.e., the symmetry before extending by $U(1)$.

 The subalgebra of (\ref{Auaral}) 
corresponding to integer, positive powers of $x,p$, denoted in the 
literature by  $w_\infty$, 
has been considered many times and very recently in connection with the 
Geroch group \cite{Geroch}. The traditional restriction to integer,
positive indices is based on analyticity grounds. However, applied to the
quantum world, the analyticity requirement makes sense for only those operators
which are not basic, i.e. are not directly associated with any degree of 
freedom and must accordingly be written in terms of the basic quantum operators
(${\hat{\os}^*},\hat{\os}$ in our case). 
Conversely, Poisson algebra elements that generate Lie algebra
cohomology (and, therefore, central extensions) can be kept as generators of 
the true quantum symmetry, as they do not have to be expressed,
in principle, as fucntions depending on ${\hat{\os}^*},\hat{\os}$. They 
will be referred to as an  ``essential anomaly" and extend 
the system in the sense that they generate new (independent) quantum 
degrees of freedom. Only the presence of anomalies will require a 
further reduction of
the quantum representation, which is achieved in a way that permits some 
{\it a priori} basic operators to be written in terms of others 
effectively basic. 
The quantum values of the anomalies are in general those values 
of the central charges for which the effective extent of the extended 
system reduces to a minimum. In any case, and as a minor harm, if we 
wish to put the motivation (\ref{motivation}) to the algebra (\ref{Auaral}) 
in a proper mathematical ground,  we could just 
eliminate the point $\os=\osc=0$ of our original phase space thus restoring
the analyticity of the combinations (\ref{motivation}) [note also 
that the quantum 
analogue $\widehat{\osc\os}=\ac\a+\um$ of the classical function 
$\osc\os$ is never zero because of the anomalous value ($\not= 0$) 
of the symplin (zero-point energy)].

To understand fully the interplay among (a certain degree of) classical 
non-analyticity, group cohomology and the extent of a quantum system, let 
us restrict ourselves to the unextended ``string" algebra (\ref{string}).
The generators of the classical algebra of symmetries are written as 
non-analytical 
functions [in the ``weak'' (avoidable) sense specified in the previous 
paragraph]  
of ${\osc},\os$:
\bea
\A_n=\um \os^{2n}({\osc}\os)^{1-n}\;\;&,& \A_{-n}=\um {\osc}^{2n}
         ({\osc}\os)^{1-n}\nn\\
\alpha_m=\um \os^{2m}({\osc}\os)^{-m}\;\;\;&,& \alpha_{-m}=
        \um {\osc}^{2m}({\osc}\os)^{-m}\;.
\eea
\ni Centrally extending this algebra in the form:
\bea
\left[\hat{\A}_n,\hat{\A}_m\right]&=&(n-m)\hat{\A}_{n+m}+
           \frac{1}{12}(cn^3+c'n)\delta_{n+m,0}\hat{1} \\
\left[\hat{\A}_n,\hat{\alpha}_m\right]&=&\hat{\alpha}_{n+m} \nn\\
\left[\hat{\alpha}_n,\hat{\alpha}_m\right]&=&am\delta_{n+m}\hat{1}\;
,\label{extension}
\eea
\ni we can proceed with group quantization, finding the characteristic 
subalgebra as well as the canonically conjugate pairs. 
The precise calculations can be found in \cite{Virasoro} and references 
therein (for the actual string algebra, i.e. for generators $\A_n,
\alpha_m^{\mu}, \mu=0,1,2,...d$, although the results are formally equivalent).
We arrive at the results given in Sec. 3: for $a=1, c=c'=1$, the whole set of 
Virasoro generators can be expressed, after quantization, as quadratic 
(hence analytical) functions of the quantum operators $\hat{\alpha}_m$
(see (\ref{Sugawara})). These operators, however, need not be (nor indeed 
can be) expressed in terms of any operator since they are basic, independent
operators, as a consequence of the central extension (\ref{extension}) 
(the central term 
in the Virasoro commutator is due to an anomaly, which is destroyed for
the values of $c,c',a$ above), giving an infinite 
extent to the physical system.
The same clearly applies to the case $\hs=-\frac{d}{4}$
of the elementary  particle with symplin studied in 
Sec. 3 where, the pseudo-extension of $SU(1,1)$ 
(which redefines the generator $\xr{\heta}$) with 
parameter $\hs$, is exactly the same as the pseudo-extension of 
the Virasoro algebra (which redefines the generator $L_0=\osc\os$) with 
parameter $c'$. 

Our suggestion, finally, is then to consider the central extensions of the 
entire (formal abstract) algebra (\ref{Auaral}),   
as being the quantizing algebra for the minimal infinite-dimensional 
system extending the free particle in such a way that string itself is 
naturally included. In this quite extended object, the free particle with 
symplin appears as the only and biggest finite-dimensional subsystem.     
Also along these lines, (1+1D) quantum gravity could arise in
a general attempt to get a full quantization of the phase space of the 
free particle. A general study of (\ref{Auaral}), its central extensions 
and quantization, will require a quite big effort and deserves a separate
work. 

\section*{Acknowledgements}

M. Calixto thanks the Spanish MEC for a FPI grant. J. Guerrero 
thanks the Dipartimento di Fisica, Universit\'a di Napoli, for its 
hospitality and the Spanish MEC for a postdoctoral grant.

\end{document}